\documentclass[10pt]{article}
\usepackage[utf8]{inputenc}
\usepackage[english]{babel}
\usepackage{cite}
\usepackage{geometry}
\usepackage{amsmath}
\usepackage{amsfonts}
\usepackage{graphicx}

\title{Information Security Games: A Survey}
\author{Erick Galinkin}
\date{\today}

\begin{document}

\maketitle

\section{Introduction}
Technology is increasingly integrated into our lives; at the same time, there are more information security-related incidents in the news. 
From Wannacry~\cite{Chen2017AutomatedRansomware} to Equifax~\cite{Wang2018CybersecurityBreach} to Solarwinds~\cite{FireEye2020HighlyBackdoor}, the implications of information security failures have become a material threat to everyday life.
As these incidents hit closer to home, the number of vulnerabilities recorded by the National Vulnerability Database~\cite{NationalInstituteofStandardsandTechnology2020NationalDashboard} continues to rise year after year, with roughly 18000 new software vulnerabilities discovered in 2020 alone. 
No one can doubt that information security is an important field of study and work, with pressing consequences that merit our attention. 
We are motivated to bring to bear any tools that can help mitigate these harms. 

As a field of study, information security integrates techniques from a variety of disciplines:
Malware detection is largely a function of information theory as developed by Shannon~\cite{Kolbitsch2009EffectiveHost}; incident response borrows techniques from forensic sciences up to and including its need for a chain of custody to ensure evidence has not been tampered with; modern cryptography is built on the foundations of number theory and computational complexity~\cite{Schneier1993AppliedCryptography, Sipser2012IntroductionComputation}. 
In addition to this sampling of applications from other fields, information security borrows extensive theory from reliability engineering~\cite{Kondakci2015AnalysisTutorial} and operations research~\cite{2017OperationsSecurity}. 
Given the applicability of disparate techniques and theories, it stands to reason that other techniques and theories could bear fruit on the rich field of information security.

Modern game theory, as developed by von Neumann and Morgenstern in their 1944 book~\cite{vonNeumann1944TheoryBehavior} has been a staple of economic analysis for decades. 
Game theory has also been applied in other contexts, including clinical psychology training~\cite{Cornell2018TrainingTraining} and national critical infrastructure security~\cite{Brown2006DefendingInfrastructure}. 
A noteworthy comment from von Neumann in~\cite{vonNeumann1944TheoryBehavior} is that in the context of economic utility, ``One of the chief difficulties lies in properly describing the assumptions that have to be made about the motives of the individual''. 
This comment could just as easily be made about the investigation of any information security incident, and so a similar framing applies to the issue of establishing an optimal defensive posture against motivated unknown adversaries.

\section{Preliminaries}
Before we explore the literature and consider the interplay between information security and game theory, we establish some definitions used throughout this document.
Some of these game conditions are not stated explicitly in the literature, so we map particular game conditions to their respective information security references in Section~\ref{sec:infosecgames}.

\subsection{Information Security} \label{sec:infosec}
We use a standard definition of information security~\cite{Harris2018CISSPGuide} as the protection of the confidentiality, integrity, and availability of information.
In this definition, confidentiality is the principle that information is only available to be read by authorized entities; integrity is the principle that information is only available to be created or modified by authorized entities; and availability is the principle that authorized entities should be able to read or write information when they need to.
Attackers seek to subvert these principles for a variety of reasons including economic gain, political motivations, intellectual property theft, and others.
Defenders seek to prevent initial access by attackers whenever able; if they fail, they seek to detect and eliminate the attacker as quickly as possible.

\subsubsection{Tactics, Techniques, and Procedures of Attackers}
Though the motivation of an attacker may be unknown, the stages of an attack at a high level are:
\begin{enumerate}
    \item \textbf{Access}: get onto the target
    \item \textbf{Persistence}: stay on the target
    \item \textbf{Action}: accomplish the objective
\end{enumerate}
There are far more in-depth models of attacker behavior from Shostack~\cite{Shostack2014ThreatModeling}, the Cyber Kill-Chain~\cite{LockheedMartinCyberHttps://www.lockheedmartin.com/en-us/capabilities/cyber/cyber-kill-chain.html}, or MITRE ATT\&CK~\cite{MITREMITREHttps://attack.mitre.org}, but the above three stages roughly cover these more granular models.

Attackers leverage a variety of methods to gain access to a potential victim.
These methods include phishing, the sending of emails crafted to entice victims to share information or inadvertently allow execution of malicious code; exploitation of vulnerabilities in software, which can allow arbitrary remote code execution on a victim machine; and using legitimate credentials that have been leaked or stolen.

The second stage of attack is persistence.
This is accomplished through some type of backdoor access, which can take a number of forms including remote access tools, or legitimate accounts that are intended for later access to a target system.
The idea behind persistence is that an adversary may want to maintain access to a system or network over a period of time.
Attackers may lay in wait, checking regularly on their persistence mechanism, or they may use a persistence method that regularly sends back status information.
One other common tactic for maintaining persistence is lateral movement.
Lateral movement is the compromise and gaining of persistence on additional machines on a target network - moving laterally to ensure that if the attacker is detected on one machine, they have other machines within the same network to reassert themselves.

Once an attacker has persistent access to a target, they can take action to accomplish their objective. 
This typically means at least one of the following:
\begin{enumerate}
    \item Steal information from the victim, known as exfiltration.
    \item Install some sort of crimeware, a type of malicious software designed for making a profit off of a victim. Typically this is either a coin miner for generating cryptocurrency, or software that encrypts files until a ransom is paid, known as ransomware.
    \item Use the victim to launch additional attacks, using \textit{e.g.}, a botnet, which further obfuscates the source of the attack.
\end{enumerate}

\subsubsection{Defending Systems}
On the other side, defenders seek to ensure that the confidentiality, integrity, and availability of networks is maintained. 
This is accomplished through the deployment of sensors: firewalls and network intrusion detection systems to detect and mitigate unauthorized network access; endpoint detection and antivirus systems to mitigate unauthorized behavior on the endpoint; and security information and event management (SIEM) as well as security orchestration, automation, and response (SOAR) software to aggregate the security data.
Taken together, these solutions form the sensors available to defenders to see within their network in order to mitigate, detect, and prevent security incidents.

Defenders are often not in control of what servers and computers are deployed in their environment, since a different team in their organization will set up and deploy servers.
Typically - a corporate information security team will be permitted to install endpoint protection software and will have a modest budget to deploy and manage their network security devices. 
The challenge that defenders must overcome then, is quite stark.
In the face of limited resources and one or more potential adversaries with unknown capability and unknown objectives, defenders must try to ensure the confidentiality, integrity, and availability of computer systems for their users.

\subsection{Game Theory} \label{sec:gametheory}
Game theory is the study of strategic decision making among rational actors, where a game, $\Gamma$ is comprised of 3 things: the set of players, $i \in I$, the set of pure strategies for each player, $\sigma_i \in \Sigma_i$, and the set of payoff or utility functions $u_i(\sigma_i)$ for each strategy.
Games have a variety of characteristics that dictate how equilibria can be found.
Zero sum games are games in which the utility gained by one player is exactly the loss encountered by the other player. 
By contrast, general sum games are games in which the utility for both players may be both positive, both negative, or one positive and the other negative. 
Games may be simultaneous, where both players decide their move and then move at the same time, or sequential, where players take turns making their moves.

Additionally, games are modeled with information about strategies and state that may or may not be available to one or more players.
In a complete information game, all information about players' payoff functions, objectives, and available strategies are available to all players. 
Incomplete information games, by contrast, are games where a player's objective may be unknown.
In a security context, this equates to games where the attacker's objective and strategies are known -- \textit{e.g.}, stealing private information about employees using a remote access tool -- and the defender's available strategies are known -- \textit{i.e.}, what capability defenders have to detect and mitigate an attack.
In a perfect information game, all information about the state of the game is available to all players. 
In imperfect information games, some information about the game state is hidden.
In a security context, perfect information equates to games where the attacker's position is known -- \textit{e.g.}, the defenders know the nodes on their network that have been infected and the attacker's source IP range.
By contrast, a security game with imperfect information would have limited visibility, where attackers do not know the full network topology of the target network and defenders are uncertain about where in their network attackers have access.

The Prisoner's Dilemma~\cite{vonNeumann1944TheoryBehavior} is the canonical example of a simultaneous move game with complete and perfect information. 
Both players know exactly the payoff functions and current state of affairs but must make their decision at the same time about whether to confess to the crime or betray the other prisoner.
This differs from Chess, which is a sequential game of perfect, complete information where all players know the objective - to checkmate the opponent's king - and no pieces on the board are hidden from either player.
We can also consider poker, which is a game of complete but imperfect information.
We know what each player's payoff is as a function of the money that has been bet, but we do not know what cards are in each player's hand.
The board game Ticket to Ride is an example of incomplete but perfect information.
In this game, we know all of the resources available to each player and all of their moves, but their win condition is hidden information.
Finally, Bayesian games are games of incomplete, imperfect information where players payoff functions are unknown and so are modeled using a probability distribution representing a ``belief'' about what a payoff will be. 
These games are detailed further in the context of information security games in Section~\ref{sec:motivation}

\subsubsection{Nash Equilibria}
The central conceit of the Nash equilibrium is that there is no unilateral change in strategy any player can make to improve their payoff.
To understand this in depth, we introduce the concept of pure and mixed strategies. 
A strategy in this context is an algorithm for playing the game: the set of all moves a player will take in the course of playing the game.
In a pure strategy, the strategy and the moves are decided and our strategy is a set of moves $\sigma = \{s_1, s_2, ..., s_n\}$ for each of the n turns of the game.
In a single move, simultaneous game like the Prisoner's Dilemma or rock-paper-scissors, a pure strategy $\sigma$ is exactly the move we will make.
Given our set of utility functions $U$, players $I$, and pure strategies $\Sigma$, we say a strategy profile $\sigma^{*}$ is in Nash Equilibrium for each player $i \in I$ if:
\begin{equation}
    u_{i}(\sigma^{*}_{i}, \sigma^{*}_{-i}) \geq u_{i}(\sigma_i, \sigma^{*}_{-i}), \forall \sigma_i \in \Sigma_i
\end{equation}
Where $\sigma^{*}_{-i}$ is the optimal strategy for players other than player $i$.

A mixed strategy $M$ is a probability density over the set of pure strategies, which allows a player to play each pure strategy with some probability.
In a mixed strategy, we can estimate expected payoff based on the probability of selecting a particular strategy and the payoff of that pure strategy.
In his landmark paper~\cite{Nash1951Non-CooperativeGames}, Nash demonstrated that a finite non-cooperative game always has at least one mixed-strategy equilibrium point.
Then, given utility functions $U$, players $I$, and mixed strategies $M$, a mixed-strategy profile $m^{*}$ is in Nash Equilibrium for each player $i \in I$ if:
\begin{equation}
    \overline{u}_i(m^{*}_{i},m^{*}_{-i}) \geq  \overline{u}_i(m_i, m^{*}_{-i}), \forall m_i \in M
\end{equation}
where $\overline{u}_i$ is the expected payoff function.

\section{Information Security Games} \label{sec:motivation}
We consider information security games generally before discussing the body of literature in Section~\ref{sec:infosecgames}.
Given our definition of a game in Section~\ref{sec:gametheory}, we define an information security game, $\Gamma$.
Though it is possible to have multiple attackers or a coalition of defenders, all information security games will have - at least - an attacker, $a$ and a defender, $d$ so $I = \{a, d\}$.
Since the defenders seek to defend a network of computers and the attackers seek to accomplish objectives on target machines, we also define our environment $\Omega = \{\omega_{g}\}, \quad\forall g \in G$, where $G$ is our network and each network node $\omega_{g}$ can take on a finite number of states, \textit{e.g.}, compromised, offline, clean.

The attacker and defender each have a set of pure strategies such that our set of strategies is defined as $\Sigma = \{\Sigma_{a}, \Sigma_{d}\}$.
Then, each player has a utility function corresponding to the value derived from the strategy, so our set of utility functions is $U = \{U_{a}, U_{d}\}$, the set of utility functions for the attacker and the defender. 
We define $U_{a}(\omega, \sigma): \Omega \times \Sigma_{a} \rightarrow \mathbb{R}$ and $U_{d}(\omega, \sigma): \Omega \times \Sigma_{d} \rightarrow \mathbb{R}$ to be the utility functions for the attacker and defender respectively.
That is, the utility is the quantifiable value for following the strategy $\sigma \in \Sigma$ given the state $\omega \in \Omega$.
Then, a general information security game is the 4-tuple:
$$\Gamma = (I, \Sigma, \Omega, U)$$

\begin{figure}
    \centering
    \includegraphics[width=\textwidth]{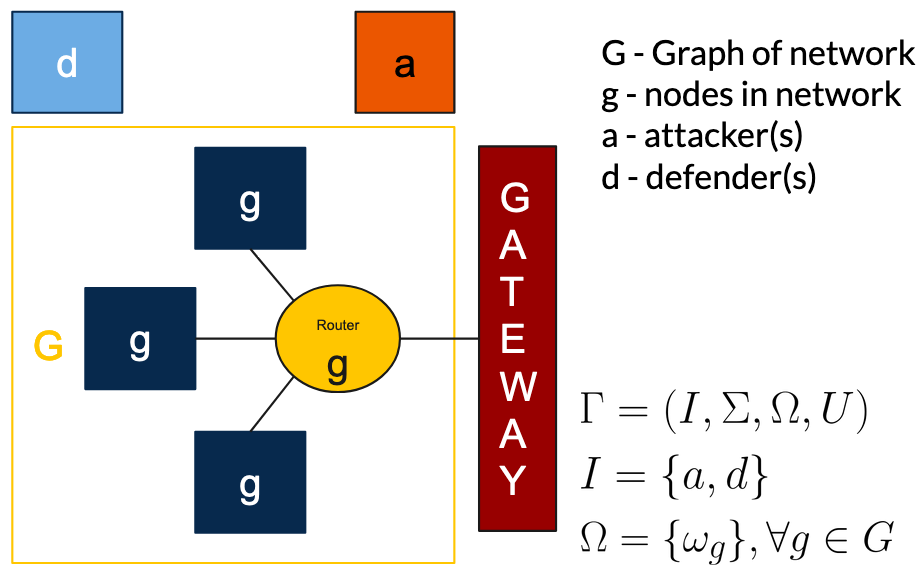}
    \caption{Example network representing the game state}
    \label{fig:example_game}
\end{figure}

This naturally raises questions about the value of the utility functions, the possible moves in the strategy set, and so on.
These questions have a variety of possible answers that require varying assumptions.
In zero-sum games, authors assume that $U_{a} = -U_{d}$, while general sum games allow for other formulations.
In complete information games, the defender knows $\Sigma_{a}$, the set of strategies available to the attacker, while others assume that this is unknown to the defender.
Information security games with perfect information assume that all of the values in $\Omega$ are public -- both the attacker and defender know all nodes that are and are not compromised -- while other researchers assume that this is mostly hidden information.

Figure~\ref{fig:example_game} contains a simple example of how a security scenario can be described as a game. 
Assume a zero-sum, simultaneous game, using the definitions in Section~\ref{sec:gametheory}.
For example, in the scenario where the attacker $a$ seeks to execute a denial of service attack against the network $G$ according to some strategy $\sigma_a$.
The nodes of the graph $G$ are three endpoints and a router behind a gateway.
Each node $g$ has a state $\omega_g$ corresponding to ``online'' or ``offline''.
The defender $d$ seeks to ensure that $\omega_g$ is ``online'' for each $g \in G$ by choosing the strategy $\sigma_d$.
If any node in $G$ is offline, then $a$'s attack is successful and they win the game.
Then, our payoffs for the defender are 1 utility unit if the nodes all remain online and -1 utility unit if any node goes offline.
Correspondingly, the attacker gets 1 utility unit if any node goes offline, and -1 utility unit if all nodes stay up, due to the cost of conducting the attack.

\section{Literature Review}\label{sec:infosecgames}
The study of games as it relates to the security of information systems originates with Syverson~\cite{Syverson1997AComputation} and the role of game theory was expanded in Hamilton \textit{et al.}~\cite{Hamilton2002TheWarfare}. 
Brown \textit{et al.}~\cite{Brown2006DefendingInfrastructure} build on this, actually going through the mathematics and modeling the possibility of attack on cyber-physical systems as both a simultaneous game and a sequential game. 
The authors determine that in both simultaneous games and games where attackers move first, the attackers are at a substantial advantage.
This is the same conclusion reached by He \textit{et al.}~\cite{He2008AAssessment}, who strictly consider attacks on computer networks. 
He \textit{et al.} use a game theoretical attack defense model and conclude that sequential, perfect information games, leader-follower\footnote{Leader-follower games are often referred to as Stackelberg games in the literature, after Heinrich Freiherr von Stackelberg (1905-1946).} games~\cite{Fudenberg1991GameTheory}, which are a more appropriate model than simultaneous games for use in security contexts.

Despite the prevailing wisdom that security games are not general sum, since there are obvious conditions under which the cost to the attacker and the cost to the defender are both too high to be recouped by any utility gained, some authors assume that the game is zero-sum, which allows them to use powerful optimization methods.
Khouzani \textit{et al.}~\cite{Khouzani2012Saddle-pointAttack, Khouzani2011AAttack} make this assumption to leverage a saddle-point strategy for finding equilibria.
This model relies on a dynamical systems view of a worm - a form of self-propagating malware - based on epidemic modeling. 
Using this dynamical systems view, under the assumption that the utility gained by the attacker is exactly the cost to the defender, Khouzani \textit{et al.} are able to deterministically compute the equilibrium for the leader-follower security game.
Since the only zero-sum games considered are those of Khouzani \textit{et al.}~\cite{Khouzani2011AAttack}, we categorize the remaining literature into a matrix by its information perfection and completion in Table~\ref{tab:citations}.

The conditions of complete and incomplete information add another layer of decision making to modeling our security games.
As noted in Section~\ref{sec:gametheory}, complete information implies that all players' strategies and payoffs are known to all other players. 
This is a common assumption in the literature, since having incomplete information about payoffs significantly changes the dynamics of the games.
Given the variety of possible motivations for cyber attack, determining the attacker's objective can be a difficult hurdle to overcome.
Chatterjee~\cite{Chatterjee2015QuantifyingPayoffs, Chatterjee2016PropagatingAnalysis} extensively considers how to deal with incomplete information in games.
In Chatterjee \textit{et al}.'s work on quantifying uncertainties in attacker payoffs~\cite{Chatterjee2015QuantifyingPayoffs}, the systemic uncertainty about attacker payoffs - that we cannot know for sure what an attacker's motivations and objectives are - are modeled as a probability distribution.
This provides a useful framework since not only will different attackers have different payoffs, but the tactics, techniques, and procedures used by the same adversary against different targets may differ.
In later work on uncertainty propogation~\cite{Chatterjee2016PropagatingAnalysis}, they consider how to propogate this uncertainty about payoffs in a multi-stage, sequential game with incomplete information.

Perfect information, as described in Section~\ref{sec:gametheory}, implies that there is no hidden information about the game state.
In security terms, this approximately means that a defender knows what machines have been compromised by an adversary, what tools they are using, and so on. 
Perfect information is not a very common assumption in the literature, but is seen in Luo \textit{et al}.'s work using fictitious play~\cite{Luo2010GameSecurity} wherein the ``administrator'' does not know what the attacker is after - thereby having perfect but incomplete information - but seeks to predict the attacker's next action and stop their advance. 
Overwhelmingly, the literature expects that we have imperfect information, seen most evidently in stochastic games where we have complete but imperfect information~\cite{Shapley1953StochasticGames}. 
In this case, the imperfection in our information stems from uncertainty about what the effect of a particular action will be: we cannot deterministically know the state of the environment given an action.
All stochastic games~\cite{Williamson2012ActiveGames, Cui2008ASystem, Sallhammar2006OnEvaluation, Nguyen2009StochasticNodes, Bommannavar2011SecurityLearning} include the incomplete information assumption that there is some uncertainty about the game state.
In the papers by Williamson \textit{et al.}~\cite{Williamson2012ActiveGames} and Bommannavar \textit{et al.}~\cite{Bommannavar2011SecurityLearning}, security games are the opening salvo in attacking the problem of using reinforcement learning in security.
By framing the problems in a game theoretic context, both papers demonstrate how policies can be learned based on malware behavior to cluster malware and manage risk, respectively.

In the aforementioned works, only 2-player games are considered. 
In n-person games, where $n > 2$, we must factor in that some subset of the players may cooperate.
When we have a subset of players cooperating to achieve their objectives, we call that subset a coalition~\cite{Myerson1997GameTheory}.
In the security context, this should be a common consideration, as there exist many independent groups of attackers and many groups within a company who may all have differing strategies for attack and defense, respectively.
Saad \textit{et al.}~\cite{Saad2010CoalitionalManagement} consider a real-world organization where there are many players, both attackers and defenders, and how to factor-in the cooperation among those players in security games. 
They consider risk minimization in an organization with many stakeholders in both the ideal communication scenario, where there is no cost to communication; and with cost or ``friction'' of communication among defender groups.
Although this is an important factor in real-world decision making, it is rarely considered in the literature.

\begin{table}[]
\begin{center}
\begin{tabular}{l|l|l}
& \begin{tabular}[c]{@{}l@{}}\textbf{Perfect} \\ \textbf{Information}\end{tabular} & \begin{tabular}[c]{@{}l@{}}\textbf{Imperfect} \\ \textbf{Information}\end{tabular}                             \\ \hline
\begin{tabular}[c]{@{}l@{}}\textbf{Complete}\\ \textbf{Information}\end{tabular}   & \begin{tabular}[c]{@{}l@{}}He~\cite{He2008AAssessment}\\ Khouzani~\cite{Khouzani2011AAttack}\\ Saad~\cite{Saad2010CoalitionalManagement}\end{tabular}   & \begin{tabular}[c]{@{}l@{}}Bommannavar~\cite{Bommannavar2011SecurityLearning}\\ Cui~\cite{Cui2008ASystem}\\ Nguyen~\cite{Nguyen2009StochasticNodes}\\ Sallhammer~\cite{Sallhammar2006OnEvaluation}\\ Williamson~\cite{Williamson2012ActiveGames}\end{tabular} \\ \hline
\begin{tabular}[c]{@{}l@{}}\textbf{Incomplete}\\ \textbf{Information}\end{tabular} & Luo~\cite{Luo2010GameSecurity} & \begin{tabular}[c]{@{}l@{}}Chatterjee~\cite{Chatterjee2015QuantifyingPayoffs}\\ Liu~\cite{Liu2006ANetworks} \\ Sartea~\cite{Sartea2020BayesianAnalysis}\end{tabular}                                    
\end{tabular} 
\caption{\label{tab:citations} Matrix of literature by Information Perfection/Completion}
\end{center}
\end{table}

The theory of mechanism design~\cite{Nisan2001AlgorithmicDesign} enables researchers to analyze the results of these games to develop security mechanisms for broader use.
This can entail balancing risk, changing incentives for attackers, or allocating finite security resources optimally.
Additionally, games can provide a framework for designing protocols that enforce truthfulness~\cite{Micali2008ResilientAuctions}.

\section{Themes and Gaps}
Broadly, the literature is reflective of an important theme in security writ large: uncertainty is everywhere.
It is rare that we know what an attacker's objective is, and correspondingly, it is difficult to model their payoff functions.
As a result, we see that in the majority of the literature - and especially in the most recent literature - we must account for that uncertainty by having imperfect information in our games.
However, researchers do not agree on whether or not information security games are best modeled as complete information games or incomplete information games, with a recent trend of using incomplete information in general and Bayesian games in particular. 
Theoretically, we can have complete information about the state of our network using various defender visibility tools as outlined in Section~\ref{sec:infosec}.
In practicality, it is difficult to get truly complete information about an environment since detection of malicious activity is formally undecidable~\cite{Cohen1985ComputerViruses} though potentially unwanted behavior can be detected and mitigated~\cite{Kolbitsch2009EffectiveHost}.

As a rule, the battlefield on which the war is assumed to be waged is the network and the endpoint is ignored in the majority of the relevant literature. 
The exception to this rule is Sartea~\cite{Sartea2020BayesianAnalysis}, who focuses on identifying the malware's type using game theory, but not preventing malware execution.
Not only are attacker techniques on the endpoint preventable with common security software, but adversaries will often stay on an endpoint for extended periods of time before moving to another node on the network~\cite{Steffens2020AttributionThreats}, eventually infecting many nodes before accomplishing their objective.
Additionally, widespread adoption of cloud infrastructure has dramatically changed the landscape from the model of a single corporate network with a unified perimeter that encloses individual servers and workstations to a broad and diverse collection of traditional networks, software as a service, cloud infrastructure, and remote workers. 
This cloud security perspective is also largely missing from the body of literature, with a single paper by Furuncu \textit{et al}.~\cite{Furuncu2015ScalableCCRAM}, which considers a simultaneous move risk-minimization game, forming the only significant work on this topic.

All prior work we uncovered concentrates on the perspective of the defender, assuming that an adversary has static capabilities and that defenders first prepare and then respond to attacks. 
As a result, the current literature fails to portray the reality of attacker payoffs and fails to identify them as regretful or no regret learners~\cite{Roughgarden2016TwentyTheory}.
This forms a noteworthy gap, as attackers are often quite innovative and have been known to leverage a variety of techniques, up to and including using authorized infrastructure such as cloud service providers~\cite{Galinkin2019TheCloud} to conduct their operations. 
In colloquial information security parlance, this would require adopting a ``red team mindset'' about security games, focusing on what attackers are motivated by and what their utility is, concretely.
Further, we did not uncover any game theoretic case studies of high profile attacks, which could also help provide this important insight.

The literature contained a number of rigorous mathematical models and assumptions. 
Each model was then used to find a Nash equilibrium given the assumptions at hand.
In the field of information security, assumptions and models can be discarded quite quickly if they are not built on empirical data and evaluated using real-world scenarios~\cite{Shostack2014ThreatModeling, Tambe2012ComputationalChallenges}.
This leaves a considerable gap in the literature for real-world evaluation of models along the lines of MITRE's CALDERA~\cite{MITRE2019TestingAdversaries}, using specific attacker simulations to assess the efficacy of a model.
Critically -- the network security game literature reviewed does not design mechanisms on top of their games~\cite{Micali2008ResilientAuctions} but instead stops upon finding a Nash equilibrium for each scenario.
In a small handful of cases, the models are used to solve relatively easy problems, such as malware family classification~\cite{Sartea2020BayesianAnalysis, Williamson2012ActiveGames}.

The body of literature also considers largely only 2-person games: games with one attacker and one defender.
With the exception of Saad \textit{et al}.~\cite{Saad2010CoalitionalManagement}, coalitional game theory was absent.
This is noteworthy, as security systems and personnel are not usually tightly integrated in practice. 
This means that security teams seeking to mitigate attacker success will nearly always need to cooperate with other groups in their organization such as end users and system administrators.
Moreover, the systems at play are not independent, nor are they monolithic.
Despite this, no discussion of compositionality in games~\cite{Ghani2018CompositionalTheory} was uncovered in the literature.

\section{Conclusion and Future Work}
The field of information security games is nascent but with considerable potential to become a powerful analytical tool researchers can use to design products and protocols to protect modern computing systems infrastructure.
By conducting equilibrium analysis, defenders are able to measure their risk in a meaningful and quantitative way.
This makes game theory a valuable tool even with no broad consensus on how modeling should be performed.
Existing approaches show that there is value in modeling the tension between attackers and defenders as a non-cooperative game, including as a coalitional game. 
A number of notable gaps remain which provide avenues for future work.

\begin{enumerate}
    \item \textbf{Compositionality} - The varied ecosystem and interplay between attacker tools and defender tools requires a compositional approach -- an approach that considers the constituent parts of the system and their interactions, rather than the system as a whole.
    The assumption of singular utility functions and game mechanisms that capture the totality of the information security game is abstracting away too much nuance.
    \item \textbf{``Red Team Mindset''} - Taking a holistic approach to modeling the game from the attacker's perspective will vastly improve the ability of future modelers to estimate accurately attacker strategies and utility.
    \item \textbf{Cloud Security Games} - Only one of the existing approaches consider cloud computing in their model, others instead assume that there is a monolithic perimeter with workstations and servers inside it.
    This is not reflective of modern computing ecosystems, which are largely hybrids of cloud and on-premises systems.
    Including cloud resources in the game model will improve the usefulness of these models by making them more realistic, even at the expense of adding complexity.
    \item \textbf{Empirical Verification} - Many of the models developed and presented throughout the literature rely on proving that a mathematical model is viable under their set of assumptions. 
    To use these models operationally, we must conduct empirical verification of our model and its assumptions with real or simulated attackers.
\end{enumerate}

\bibliography{references}

\begin{thebibliography}{10}

\bibitem{Bommannavar2011SecurityLearning}
{\sc Bommannavar, P., Alpcan, T., and Bambos, N.}
\newblock {Security risk management via dynamic games with learning}.
\newblock {\em IEEE International Conference on Communications\/} (2011).

\bibitem{Brown2006DefendingInfrastructure}
{\sc Brown, G., Carlyle, M., Salmer{\'{o}}n, J., and Wood, K.}
\newblock {Defending critical infrastructure}.
\newblock {\em Interfaces 36}, 6 (2006), 530--544.

\bibitem{Chatterjee2015QuantifyingPayoffs}
{\sc Chatterjee, S., Halappanavar, M., Tipireddy, R., and Oster, M.}
\newblock {Quantifying Mixed Uncertainties in Cyber Attacker Payoffs}.
\newblock In {\em IEEE International Symposium on Technologies for Homeland
  Security\/} (2015), IEEE.

\bibitem{Chatterjee2016PropagatingAnalysis}
{\sc Chatterjee, S., Tipireddy, R., Oster, M., and Halappanavar, M.}
\newblock {Propagating Mixed Uncertainties in Cyber Attacker Payoffs :
  Exploration of Two-Phase Monte Carlo Sampling and Probability Bounds
  Analysis}.
\newblock In {\em IEEE International Symposium on Technologies for Homeland
  Security\/} (2016), IEEE.

\bibitem{Chen2017AutomatedRansomware}
{\sc Chen, Q., and Bridges, R.~A.}
\newblock {Automated behavioral analysis of malware: A case study of wannacry
  ransomware}.
\newblock {\em Proceedings - 16th IEEE International Conference on Machine
  Learning and Applications, ICMLA 2017 2017-Decem}, 1 (2017), 454--460.

\bibitem{Cohen1985ComputerViruses}
{\sc Cohen, F.}
\newblock {\em {Computer Viruses}}.
\newblock PhD thesis, University of Southern California, 1985.

\bibitem{Cornell2018TrainingTraining}
{\sc Cornell, B.~D.}
\newblock {\em {Training Games: An Application of Game Theory to Clinical
  Psychology Graduate Training}}.
\newblock PhD thesis, University of Denver, 2018.

\bibitem{Cui2008ASystem}
{\sc Cui, X., Tan, X., Zhang, Y., and Xi, H.}
\newblock {A markov game theory-based risk assessment model for network
  information system}.
\newblock {\em Proceedings - International Conference on Computer Science and
  Software Engineering, CSSE 2008 3}, 2006 (2008), 1057--1061.

\bibitem{2017OperationsSecurity}
{\sc Daras, N., and Rassias, T.}, Eds.
\newblock {\em {Operations Research, Engineering, and Cyber Security}}.
\newblock Springer, 2017.

\bibitem{FireEye2020HighlyBackdoor}
{\sc {FireEye}}.
\newblock {Highly Evasive Attacker Leverages SolarWinds Supply Chain to
  Compromise Multiple Global Victims with SUNBURST Backdoor}, 2020.

\bibitem{Fudenberg1991GameTheory}
{\sc Fudenberg, D., and Tirole, J.}
\newblock {\em {Game Theory}}, first~ed.
\newblock The MIT Press, 1991.

\bibitem{Furuncu2015ScalableCCRAM}
{\sc Furuncu, E., and Sogukpinar, I.}
\newblock {Scalable risk assessment method for cloud computing using game
  theory (CCRAM)}.
\newblock {\em Computer Standards and Interfaces 38\/} (2015), 44--50.

\bibitem{Galinkin2019TheCloud}
{\sc Galinkin, E., Singh, A., Vamshi, A., Hwong, J., Estep, C., and Canzanese,
  R.}
\newblock {The Future of Cyber Attacks and Defense is in the Cloud}.
\newblock In {\em Proceedings - IEEE MALCON\/} (2019).

\bibitem{Ghani2018CompositionalTheory}
{\sc Ghani, N., Hedges, J., Winschel, V., and Zahn, P.}
\newblock {Compositional game theory}.
\newblock In {\em LICS '18: Proceedings of the 33rd Annual ACM/IEEE Symposium
  on Logic in Computer Science\/} (2018), pp.~472--481.

\bibitem{Hamilton2002TheWarfare}
{\sc Hamilton, S., Miller, W., Ott, A., and Saydjari, O.}
\newblock {The role of game theory in information warfare}.
\newblock {\em 4th Information Survivability Workshop\/} (2002), 1--4.

\bibitem{Harris2018CISSPGuide}
{\sc Harris, S., and Maymi, F.}
\newblock {\em {CISSP All-in-One Exam Guide}}, eighth~ed.
\newblock McGraw-Hill, 2018.

\bibitem{He2008AAssessment}
{\sc He, W., Chunhe, X., Wang, H., Zhang, C., and Ji, Y.}
\newblock {A game theoretical attack-defense model oriented to network security
  risk assessment}.
\newblock {\em Proceedings - International Conference on Computer Science and
  Software Engineering, CSSE 2008 3\/} (2008), 498--504.

\bibitem{Khouzani2011AAttack}
{\sc Khouzani, M.~H., Sarkar, S., and Altman, E.}
\newblock {A dynamic game solution to malware attack}.
\newblock {\em Proceedings - IEEE INFOCOM\/} (2011), 2138--2146.

\bibitem{Khouzani2012Saddle-pointAttack}
{\sc Khouzani, M.~H., Sarkar, S., and Altman, E.}
\newblock {Saddle-point strategies in malware attack}.
\newblock {\em IEEE Journal on Selected Areas in Communications 30}, 1 (2012),
  31--43.

\bibitem{Kolbitsch2009EffectiveHost}
{\sc Kolbitsch, C., Comparetti, P.~M., Kruegel, C., Kirda, E., Zhou, X., and
  Wang, X.~F.}
\newblock {Effective and efficient malware detection at the end host}.
\newblock {\em Proceedings of the 18th USENIX Security Symposium\/} (2009),
  351--366.

\bibitem{Kondakci2015AnalysisTutorial}
{\sc Kondakci, S.}
\newblock {Analysis of information security reliability: A tutorial}.
\newblock {\em Reliability Engineering and System Safety 133\/} (1 2015),
  275--299.

\bibitem{Liu2006ANetworks}
{\sc Liu, Y., Comaniciu, C., and Man, H.}
\newblock {A Bayesian game approach for intrusion detection in wireless ad hoc
  networks}.
\newblock {\em ACM International Conference Proceeding Series 199}, January
  2006 (2006).

\bibitem{LockheedMartinCyberHttps://www.lockheedmartin.com/en-us/capabilities/cyber/cyber-kill-chain.html}
{\sc {Lockheed Martin}}.
\newblock {Cyber Kill-Chain.
  https://www.lockheedmartin.com/en-us/capabilities/cyber/cyber-kill-chain.html}.

\bibitem{Luo2010GameSecurity}
{\sc Luo, Y., Szidarovszky, F., Al-Nashif, Y., and Hariri, S.}
\newblock {Game Theory Based Network Security}.
\newblock {\em Journal of Information Security 01}, 01 (2010), 41--44.

\bibitem{Micali2008ResilientAuctions}
{\sc Micali, S., and Valiant, P.}
\newblock {Resilient Mechanisms For Truly Combinatorial Auctions}.
\newblock {\em Computer Science and Artificial Intelligence Laboratory
  Technical Report\/} (2008).

\bibitem{MITREMITREHttps://attack.mitre.org}
{\sc {MITRE}}.
\newblock {MITRE ATT{\&}CK. https://attack.mitre.org}.

\bibitem{MITRE2019TestingAdversaries}
{\sc {MITRE}}.
\newblock {Testing Your Network Defenses by Imitating Malicious Adversaries},
  2019.

\bibitem{Myerson1997GameTheory}
{\sc Myerson, R.~B.}
\newblock {\em {Game Theory}}, first~ed.
\newblock Harvard University Press, Boston, 1997.

\bibitem{Nash1951Non-CooperativeGames}
{\sc Nash, J.~F.}
\newblock {Non-Cooperative Games}.
\newblock {\em Annals of Mathematics 54}, 2 (1951), 286--295.

\bibitem{NationalInstituteofStandardsandTechnology2020NationalDashboard}
{\sc {National Institute of Standards and Technology}}.
\newblock {National Vulnerability Database Dashboard}, 2020.

\bibitem{Nguyen2009StochasticNodes}
{\sc Nguyen, K.~C., Alpcan, T., and Ba{\c{s}}ar, T.}
\newblock {Stochastic games for security in networks with interdependent
  nodes}.
\newblock {\em Proceedings of the 2009 International Conference on Game Theory
  for Networks, GameNets '09\/} (2009), 697--703.

\bibitem{Nisan2001AlgorithmicDesign}
{\sc Nisan, N., and Ronen, A.}
\newblock {Algorithmic Mechanism Design}.
\newblock {\em Games and Economic Behavior 35}, 1-2 (4 2001), 166--196.

\bibitem{Roughgarden2016TwentyTheory}
{\sc Roughgarden, T.}
\newblock {\em {Twenty Lectures on Algorithmic Game Theory}}.
\newblock Cambridge University Press, 2016.

\bibitem{Saad2010CoalitionalManagement}
{\sc Saad, W., Alpcan, T., Ba{\c{s}}ar, T., and Hj{\o}rungnes, A.}
\newblock {Coalitional game theory for security risk management}.
\newblock {\em 5th International Conference on Internet Monitoring and
  Protection, ICIMP 2010\/} (2010), 35--40.

\bibitem{Sallhammar2006OnEvaluation}
{\sc Sallhammar, K., Helvik, B.~E., and Knapskog, S.~J.}
\newblock {On stochastic modeling for integrated security and dependability
  evaluation}.
\newblock {\em Journal of Networks 1}, 5 (2006), 31--42.

\bibitem{Sartea2020BayesianAnalysis}
{\sc Sartea, R., Chalkiadakis, G., Farinelli, A., and Murari, M.}
\newblock {Bayesian active malware analysis}.
\newblock {\em Proceedings of the International Joint Conference on Autonomous
  Agents and Multiagent Systems, AAMAS 2020-May}, Aamas (2020), 1206--1214.

\bibitem{Schneier1993AppliedCryptography}
{\sc Schneier, B.}
\newblock {\em {Applied Cryptography}}.
\newblock Wiley, 1993.

\bibitem{Shapley1953StochasticGames}
{\sc Shapley, L.}
\newblock {Stochastic Games}.
\newblock {\em Proceedings of the National Academy of Sciences\/} (1953).

\bibitem{Shostack2014ThreatModeling}
{\sc Shostack, A.}
\newblock {\em {Threat Modeling}}, first~ed.
\newblock Wiley, 2014.

\bibitem{Sipser2012IntroductionComputation}
{\sc Sipser, M.}
\newblock {\em {Introduction to the Theory of Computation}}, third~ed.
\newblock Cengage Learning, 2012.

\bibitem{Steffens2020AttributionThreats}
{\sc Steffens, T.}
\newblock {\em {Attribution of Advanced Persistent Threats}}, first~ed.
\newblock Springer, 2020.

\bibitem{Syverson1997AComputation}
{\sc Syverson, P.~F.}
\newblock {A different look at secure distributed computation}.
\newblock {\em Proceedings - IEEE Computer Security Foundations Symposium\/}
  (1997), 109--115.

\bibitem{Tambe2012ComputationalChallenges}
{\sc Tambe, M., Jiang, A.~X., An, B., and Jain, M.}
\newblock {Computational game theory for security: Progress and challenges}.
\newblock In {\em AAAI 2012 Spring Symposium on Game Theory for Security,
  Sustainability and Health\/} (2012).

\bibitem{vonNeumann1944TheoryBehavior}
{\sc von Neumann, J., and Morgenstern, O.}
\newblock {\em {Theory of Games and Economic Behavior}}.
\newblock Princeton University Press, 1944.

\bibitem{Wang2018CybersecurityBreach}
{\sc Wang, P., and Johnson, C.}
\newblock {Cybersecurity Incident Handling: A Case Study of the Equifax
  Breach}.
\newblock {\em Issues in Information Systems 19}, 3 (2018), 150--159.

\bibitem{Williamson2012ActiveGames}
{\sc Williamson, S., Ong, C.~H., Williamson, S.~A., and Hui, O.~C.}
\newblock {Active Malware Analysis using Stochastic Games}.

\end{thebibliography}

\end{document}